\documentclass[10.75pt, a4paper]{article}
\usepackage{titlesec}
\titleformat{\section}
  {\bf\sffamily}
  {\thesection. }
  {5pt}
  {\MakeUppercase}
\renewcommand{\thesection}{\Roman{section}} 

\titleformat{\subsection}
  {\bf\sffamily}
  {\thesubsection. }
  {5pt}{}
  
\renewcommand{\thesubsection}{\Alph{subsection}}

\usepackage[affil-it]{authblk} 
\usepackage{etoolbox}
\usepackage{lmodern}

\usepackage{eurosym}

\usepackage[utf8]{inputenc}
\usepackage{geometry}
\usepackage[hidelinks]{hyperref}
\usepackage[citestyle=numeric,style=phys,biblabel=brackets,backend=bibtex,sorting=none,url=false,doi=false, natbib]{biblatex}
\bibliography{bibliography}
\DefineBibliographyStrings{english}{andothers={\itshape et\addabbrvspace al\adddot}}

\usepackage{csquotes}
\usepackage[]{authblk} 
\usepackage{etoolbox}
\usepackage{lmodern}

\usepackage{amsmath}
\usepackage{enumerate}
\usepackage{enumitem}
\usepackage{graphicx}
\usepackage{siunitx}
\usepackage{float}
\usepackage[]{parskip} 

\makeatletter
\patchcmd{\@maketitle}{\LARGE \@title}{\fontsize{16}{19.2}\selectfont\@title}{}{}
\makeatother

\usepackage[normalem]{ulem}

\usepackage{graphicx}
\usepackage{dcolumn}
\usepackage{bm}
\usepackage{tikz}
\usetikzlibrary{math}
\usetikzlibrary{shapes.geometric, arrows}
\usetikzlibrary{positioning}
\usetikzlibrary{shapes,arrows}
\usetikzlibrary{intersections}
\usepackage{csvsimple}
\usepackage{changepage}
\usepackage[tbtags]{mathtools}
\usepackage{siunitx}
\usepackage{csquotes}

\usepackage{float}
\usepackage{subfigure}
\usepackage[
	nonumberlist, 				
	acronym,      				
	nomain,						
	nopostdot,					
]{glossaries}
\usepackage{glossary-superragged}
\usepackage[hidelinks]{hyperref}
\usepackage{color, colortbl}

\usepackage{wrapfig}

\usepackage{pgfplots}
\usepackage{tikz}
\usetikzlibrary{arrows}
\usepackage{amsmath}
\pgfplotsset{compat=newest}
\usepgfplotslibrary{fillbetween}
\usetikzlibrary{calc}
\def\centerarc[#1](#2)(#3:#4:#5)
{ \draw[#1] ($(#2)+({#5*cos(#3)},{#5*sin(#3)})$) arc (#3:#4:#5); }

\usepackage{amssymb}

\usepackage{nicefrac}

\usepackage{abstract}

\usepackage{multirow}
\usepackage{tabularx}
\usepackage{booktabs}
\usepackage{array}
\usepackage{longtable}
\newcolumntype{L}[1]{>{\raggedright\let\newline\\\arraybackslash\hspace{0pt}}m{#1}}
\newcolumntype{C}[1]{>{\centering\let\newline\\\arraybackslash\hspace{0pt}}m{#1}}
\newcolumntype{R}[1]{>{\raggedleft\let\newline\\\arraybackslash\hspace{0pt}}m{#1}}

\newacronym{3d}{3D}{three dimensional}
\newacronym{am}{AM}{additive manufacturing}
\newacronym{fdm}{FDM}{fused deposition modeling}
\newacronym{ism}{ISM}{in-space manufacturing}
\newacronym{iss}{ISS}{International Space Station}
\newacronym{fcb}{FCB}{Functional Cargo Block}
\newacronym{dem}{DEM}{discrete element method}
\newacronym{md}{MD}{molecular dynamics}
\newacronym{dc}{DC}{direct-current}
\newacronym[plural=PFCs,firstplural=parabolic flight campaigns (PFCs)]{pfc}{PFC}{Parabolic Flight Campaign}
\newacronym{fft}{FFT}{Fast Fourrier Transform}
\newacronym{cad}{CAD}{Computer Assisted Design}
\newacronym{ptfe}{PTFE}{polytetrafluoroethylene}
\newacronym{ps}{PS}{polystyrene}
\newacronym{nasa}{NASA}{National Aeronautics and Space Administration}
\newacronym{esamm}{ESAMM}{Extended Structure Additive Manufacturing Machine}
\newacronym{amf}{AMF}{Additive Manufacturing Facility}
\newacronym{us}{US}{United States}
\newacronym{usa}{USA}{United States of America}
\newacronym{bmgs}{BMGs}{Bulk Metallic Glasses}
\newacronym{esa}{ESA}{European Space Agency}
\newacronym{si}{SI}{International System of Units, abbreviated from French \textit{Syst\`{e}me International (d'unit\'{e}s)}}
\newacronym{dlr}{DLR}{German Aerospace Center}
\newacronym{liggghts}{LIGGGHTS}{\acrshort{lammps} Improved for General Granular and Granular Heat Transfer Simulations}
\newacronym{lammps}{LAMMPS}{Large-scale Atomic/Molecular Massively Parallel Simulator}
\newacronym{sjkr}{SJKR}{Simplified Johnson-Kendall-Roberts}
\newacronym{ded}{DED}{Directed Energy Deposition}
\newacronym{slm}{SLM}{Selective Laser Melting}
\newacronym{sls}{SLS}{Selective Laser Sintering}
\newacronym{eva}{EVA}{Extra-Vehicular Activity}
\newacronym{sem}{SEM}{Scanning Electron Microscopy}
\newacronym{RPM}{RPM}{Ramdom Positioning Machine}
\newacronym{rpm}{rpm}{revolutions per minute}
\newacronym{rise}{RISE}{Research Internships in Science and Engineering}
\newacronym{daad}{DAAD}{German Academic Exchange Service, abbreviated from German \textit{Deutscher Akademischer Austauschdienst}}
\newacronym{fsm}{FSM}{finite-state machine}
\newacronym{ir}{IR}{infrared}
\newacronym{pcbs}{PCBs}{Printed Circuit Boards}
\newacronym{pcb}{PCB}{Printed Circuit Board}
\newacronym{mcr}{MCR}{Modular Compact Rheometer}
\newacronym{sff}{SFF}{Solid Freeform Fabrication}
\newacronym{uv}{UV}{ultraviolet}
\newacronym{abs}{ABS}{acrylonitrile butadiene styrene}
\newacronym{hpde}{HPDE}{high density polyethylene}
\newacronym{pei}{PEI}{polyetherimide}
\newacronym{bff}{BFF}{BioFabrication Facility}
\newacronym{lens}{LENS}{Laser Engineered Net Shaping}
\newacronym{cnc}{CNC}{Computer Numerical Control}
\newacronym{ebf3}{EBF$^3$}{Electron Beam Free-Form Fabrication}
\newacronym{leo}{LEO}{Low Earth Orbit}
\newacronym{pc}{PC}{polycarbonate}
\newacronym{crissp}{CRISSP}{Customisable Recyclable International Space Station Packaging}
\newacronym{Athena}{Athena}{Advanced Telescope for High-ENergy Astrophysics}
\newacronym{lbm}{LBM}{Laser Beam Melting}
\newacronym{bam}{BAM}{Federal Institute for Materials Research and Testing, abbreviated from German \textit{Bundesanstalt f\"{u}r Materialforschung und-pr\"{u}fung}}
\newacronym{pbf}{PBF}{powder bed fusion}
\newacronym{eb}{EB}{Electron Beam}
\newacronym{2d}{2D}{two dimensional}
\newacronym{4d}{4D}{four dimensional}
\newacronym{ft4}{FT4}{Freeman Technology 4 Powder Rheometer}
\newacronym{dsc}{DSC}{Differential Scanning Calorimetry}
\newacronym{pmma}{PMMA}{polymethylmethacrylate}
\newacronym{1g}{$1g$}{gravity on-ground}
\newacronym{mug}{$\mu g$}{microgravity}
\newacronym{bcm}{BCM}{Box Counting Method}
\newacronym{mct}{MCT}{Mode Coupling Theory}
\newacronym{gmct}{gMCT}{granular Mode Coupling Theory}
\newacronym{itt}{ITT}{Integration Through Transients}
\newacronym{mfc}{MFC}{Mass Flow Controller}
\newacronym{ct}{CT}{computed tomography}
\newacronym{xct}{XCT}{X-ray computed tomography}
\newacronym{cv}{CV}{curriculum vitae}
\newacronym{pi}{PI}{principal investigator}
\newacronym{osp}{OSP}{orthogonal superimposed perturbation}
\newacronym{npi}{NPI}{Network Partnering Initiative}
\newacronym{ecsat}{ECSAT}{European Centre for Space Applications and Telecommunications}
\newacronym{eac}{EAC}{European Astronaut Centre}
\newacronym{estec}{ESTEC}{European Space Research and Technology Centre}
\newacronym{fps}{fps}{frames per second}
\newacronym{pdf}{pdf}{probability density function}
\newacronym{al}{Al}{aluminium}
\newacronym{ss}{\textit{SS}}{\textit{Smooth Surface}}
\newacronym{rs}{\textit{RS}}{\textit{Rough Surface}}
\newacronym{rcp}{rcp}{random close packing}
\newacronym{iop}{IoP UvA}{Institute of Physics of the University of Amsterdam}
\newacronym{mp}{MP}{Institute of Material Physics for Space}
\newacronym{elgra}{ELGRA}{European Low Gravity Research Association}
\newacronym{zarm}{ZARM}{Center of Applied Space Technology and Microgravity}
\newacronym{piv}{PIV}{particle image velocimetry}
\usepackage[framemethod=tikz]{mdframed}
\usepackage{lipsum}
\usepackage{dirtytalk}
\usepackage{tcolorbox}
\usepackage[font=footnotesize]{caption}
\newtcolorbox{mybox}[1]{colback=green!6!white,colframe=black!75!black,fonttitle=\bfseries,title=#1}
\newtcolorbox{mybox2}{colback=red!5!white,colframe=red!75!black}

\usepackage{pifont}

\usepackage{soul,xcolor}
\setstcolor{red}

\usepackage{xcolor,hyperref}
\hypersetup{
   colorlinks,
   linkcolor={blue!50!black},
   citecolor={blue!50!black},
   urlcolor={blue!80!black}
} 

\definecolor{mycolor}{rgb}{0.122, 0.435, 0.698}

\title{Chloroplasts in plant cells show active glassy behavior under low light conditions.}
\author[1]{Nico Schramma\footnote{n.schramma@uva.nl, ORCID: 0000-0003-3887-3416}}
\author[1]{Cintia Perugachi Israëls}
\author[1]{Maziyar Jalaal\footnote{m.jalaal@uva.nl, ORCID: 0000-0002-5654-8505}}

\affil[1]{Van der Waals-Zeeman Institute, Institute of Physics, University of Amsterdam, Science Park 904, Amsterdam, 1098XH, The Netherlands}

\begin{document}
\definecolor{brickred}{rgb}{0.8, 0.25, 0.33}
\definecolor{darkorange}{rgb}{1.0, 0.55, 0.0}
\definecolor{persiangreen}{rgb}{0.0, 0.65, 0.58}
\definecolor{persianindigo}{rgb}{0.2, 0.07, 0.48}
\definecolor{cadet}{rgb}{0.33, 0.41, 0.47}
\definecolor{turquoisegreen}{rgb}{0.63, 0.84, 0.71}
\definecolor{sandybrown}{rgb}{0.96, 0.64, 0.38}
\definecolor{blueblue}{rgb}{0.0, 0.2, 0.6}
\definecolor{ballblue}{rgb}{0.13, 0.67, 0.8}
\definecolor{greengreen}{rgb}{0.0, 0.5, 0.0}
\begingroup
\sffamily
\maketitle
\endgroup

\begin{abstract}

Plants have developed intricate mechanisms to adapt to changing light conditions. Besides photo- and heliotropism (the differential growth towards light and the diurnal motion with respect to sunlight), chloroplast motion acts as a fast mechanism to change the intracellular structure of leaf cells. While chloroplasts move towards the sides of the plant cell to avoid strong light, they accumulate and spread out into a layer on the bottom of the cell at low light to increase the light absorption efficiency.  
Although the motion of chloroplasts has been studied for over a century, the collective organelle-motion leading to light adapting self-organized structures remains elusive. 
Here, we study the active motion of chloroplasts under dim light conditions, leading to an accumulation in a densely packed quasi-2D layer. We observe burst-like re-arrangements and show that these dynamics resemble systems close to the glass transition by tracking individual chloroplasts. Furthermore, we provide a minimal mathematical model to uncover relevant system parameters controlling the stability of the dense configuration of chloroplasts.
Our study suggests that the meta-stable caging close to the glass-transition in the chloroplast mono-layer serves a physiological relevance: chloroplasts remain in a spread-out configuration to increase the light uptake, but can easily fluidize when the activity is increased to efficiently rearrange the structure towards an avoidance state.
Our research opens new questions about the role that dynamical phase transitions could play in self-organized intracellular responses of plant cells towards environmental cues.


\textbf{keywords: chloroplast photorelocation $|$ active glasses $|$ dense active matter $|$ organelle movement} 

\end{abstract}

\begin{wrapfigure}{r}{0.5\textwidth}
\centering
\includegraphics[width=0.45\textwidth]{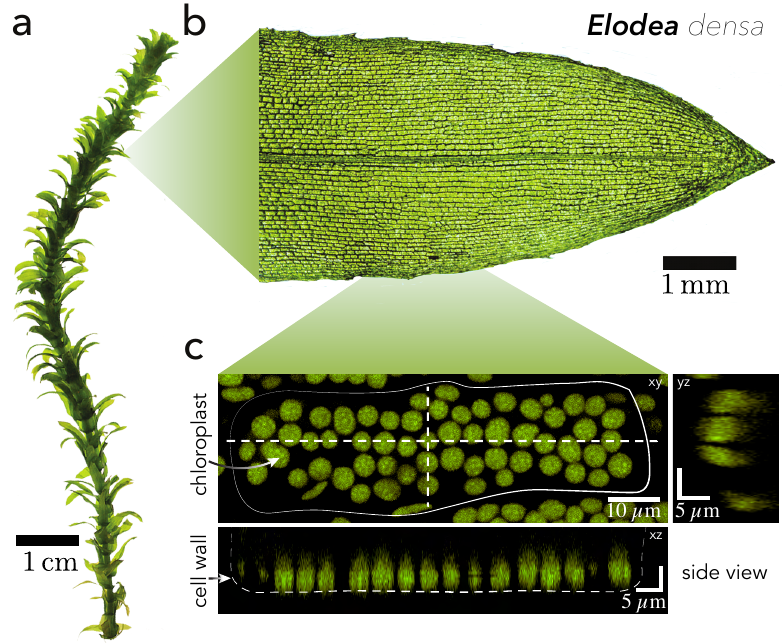}
\caption{\textbf{(a)} \textit{Elodea densa} waterplant as cultured under controlled environmental conditions. \textbf{(b)} Leaves are detached from the plants and \textbf{(c)} Laser scanning confocal microscopy image of chloroplasts inside a single epidermal cell. Chloroplasts (in green) are spread out in a single layer under dim light conditions as indicated by the side view along dotted line representing the position of the cell wall.}
\label{fig:1}
\end{wrapfigure}
Plant sensing, consciousness, and movement have astonished scientists and philosophers since ancient times~\cite{whippo2009sensational}. From Plato, who in \textit{Timaeus}~\cite{carpenter2010embodied} associated plants with a ``soul'' that lacks ``judgment and intelligence'', but shares ``sensation, pleasure, pain, and desire''; to renaissance scientists like Porta, Bacon, and Hooke who wondered about the mechanism of plant movement, and to seminal work of Charles and Francis Darwin on plant behavior~\cite{darwin1883power}. It is now known that, despite generally being limited in their mobility, plants have evolved many movement mechanisms across both length and time scales to maintain optimal development and adapt to their environment~\cite{forterre2013slow, moulton2020multiscale}.
Examples of these behavioral mechanisms are phototropism (response to light), gravitropism (response to gravity), and thigmotropism (response to touch). Such movements can occur on various time scales~\cite{forterre2013slow} and mostly on organismal length scales. 
However, the response to an environmental cue can also occur on a cellular scale. 
An example of such responses is the active reaction to light: on large scales, via phototropism, the plant grows and moves towards light~\cite{liscum2014phototropism}. Meanwhile, on a small scale, chloroplasts can rearrange to optimize the photosynthesis and respiration processes~\cite{Kasahara2002,Wada2013,Gotoh2018}. Hence, chloroplast motion enables plants to efficiently absorb light in order to generate metabolites as a product of photosynthesis while avoiding damage due to strong light.
The light-induced movement of chloroplasts was first observed in the 19\textsuperscript{th} century~\cite{Bohm1856,Frank1871}. These organelles can change the direction of movement towards or away from light~\cite{senn1908gestalts,Haupt1982} without turning~\cite{Tsuboi2009,Tsuboi2011}, but by photo-receptor dependent re-distribution of a chloroplast-specific membrane protein CHUP1~\cite{Sakai2001,Suetsugu2020,Kong2020}. This protein is essential for correct chloroplast positioning~\cite{Oikawa2003,Oikawa2008} and acts as a nucleation factor~\cite{Kong2020} for chloroplast-specific short actin filaments~\cite{Kadota2009,Kong2013}. Therefore, the CHUP1 distribution controls the binding affinity of short actin filaments which are polymerized and bundled by a plasma-membrane bound protein THRUMIN1~\cite{Whippo2011}. 
If the CHUP1 distribution is asymmetric, the binding affinity of short actin filaments is biased and leads to a net propulsion force between the chloroplast and the the plasma membrane of the cell~\cite{Wada2018}.
On the one hand, chloroplasts avoid strong light by fast escape movements~\cite{Kagawa2004, Kadota2009, Takahashi2011}, which leads to a decrease of potential photo-damage~\cite{Kasahara2002,Park1996,Li2009}. On the other hand, an accumulation response towards weak light sources~\cite{Kagawa1996,Tsuboi2011} leads to an increased absorption efficiency~\cite{Gotoh2018} as chloroplasts self-organize into a dense layer on periclinal cell walls~\cite{Zurzycki1955} (Fig.~\ref{SI_fig:SchematicBio}a). This implies that chloroplast motion simultaneously maximizes photosynthetic performance while minimizing photo-damage.\\
Chloroplasts in various plants can achieve high surface fractions of around $70\%$ upon the accumulation towards dim light~\cite{Honda1971,Ellis1985}. 
Interestingly, higher numbers of smaller chloroplasts enhances mobility~\cite{Koniger2008}, while osmotic expansion of chloroplasts leads to an effective inhibition of the light response~\cite{McCain1998}. These observations point towards a dependence of the chloroplast mobility on their packing density and their individual activity. It has been shown in other biological systems that packing density and geometry impacts the fluidity and transport properties.
Increased densities in confluent tissues can lead to a glass transition~\cite{Angelini2011,Garcia2015,Sadati2013}, arresting the cell motion and stabilizing the tissue in a \textit{solid-like} state~\cite{Atia2018,Bi2016,Kim2021}. Such rigidity transitions play a role in morphogenesis of zebrafish embryos~\cite{Mongera2018}, stabilize  bronchial epithelium of the airway during development~\cite{Park2015}, and may be key to understand cancerous tissue~\cite{Oswald2017}.
Moreover, even intracellular dynamics can exhibit features found in active granular liquids, as in the context of avalanching organelles for plant cell gravitropism~\cite{Berut2018}, or feature glass-like transport properties in the cytoplasm in bacteria~\cite{Parry2014} or human cells~\cite{Aberg2021}.
This indicates that active glassy dynamics occur across scales ~\cite{Janssen2019a} and play a key role in understanding the structural changes and transport in and around cells.\\
\indent In this study, we investigate the dynamics of chloroplast motion under dim light conditions in the water plant \textit{Elodea densa} (Fig.~\ref{fig:1}). We find that the dense, actively driven system shows dynamics similar to systems close to the glass transition, both in their overall statistics and individual particle dynamics. Despite the complexity of the system, we can construct a mathematical model that can re-produce the step size statistics with experimental and physiological parameters. We use this model to test the stability of the state close to the glass transition and show that the chloroplast motion can exhibit more \textit{liquid-like} dynamics when the activity slightly increases. 
Our observations suggest, that chloroplasts naturally organize close to a critical point in order to reach a large area coverage. Consequently, chloroplasts increase photosynthetic efficiency while maintaining the ability to quickly respond to unfavorable light conditions by actively \textit{fluidizing} the configuration for efficient avoidance movement.
\section*{Light-dependent Dynamical Phases of Chloroplasts}
\begin{figure}[htbp]
\centering
\includegraphics[width=1\textwidth]{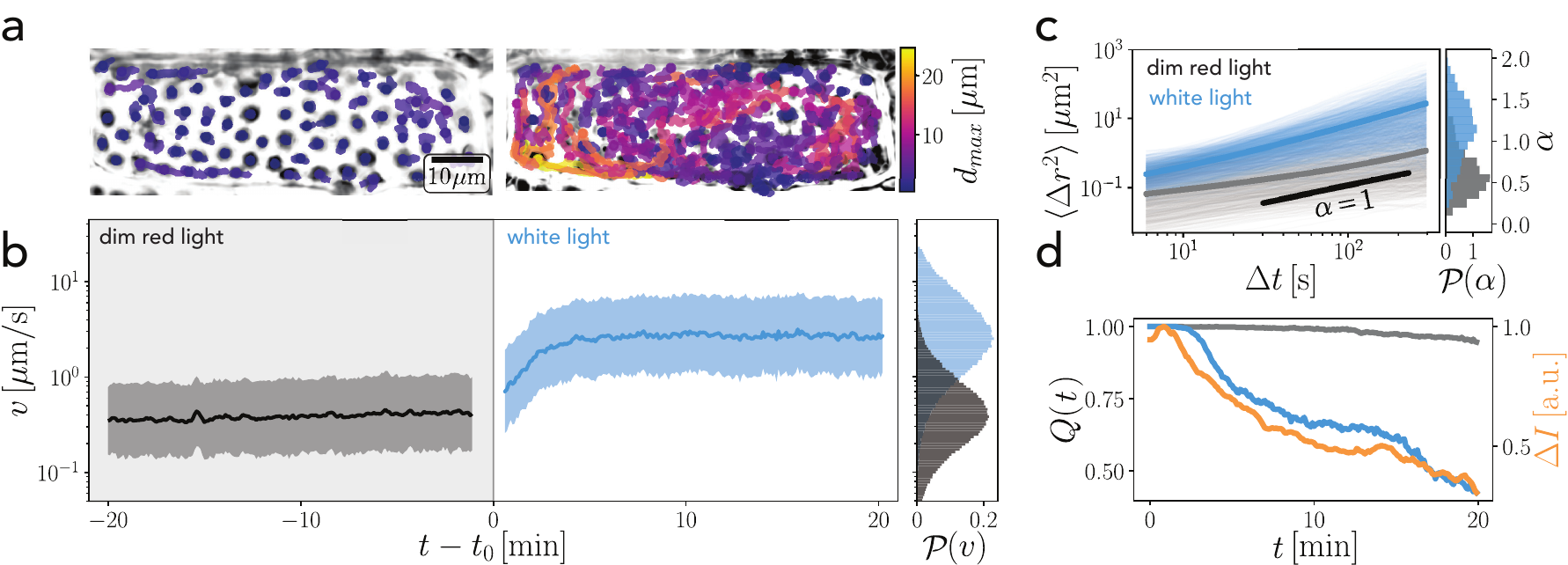}
\caption{Light adaptation of chloroplasts. \textbf{(a)} Trajectories of chloroplasts in a single cell for $20$ minutes before (left) and after (right) the onset of bright light. Color bar: maximal distance from the initial position. \textbf{(b)} The speed $v$ changes by one order of magnitude before (gray) and after (blue) the light stimulus at time $t_0$. The gap is due to the manual change of filter, light intensity and exposure time which lasted $3\,\mathrm{min}$. Right: histograms of the speed distributions for all trajectories and all times. \textbf{(c)} Mean squared displacement acts as an indicator of enhanced activity. White light (blue line) triggers mainly diffusive to super-diffusive motion at time scales around $\Delta t= 10-100\,\mathrm{s}$, while dim light adapted chloroplasts (grey) are mainly sub-diffusive. The histogram displays the probability density of anomalous diffusion exponents $\alpha$. \textbf{(d)} Self-overlap function of the dim-light adapted state (gray) and the light-adapting chloroplasts (blue). Orange line indicates the relative change in average intensity in the raw microscopy data indicating a decreasing light absorption.}
\label{fig:lightstim}
\end{figure}

We investigate the motion of $728$ chloroplasts $20\,\mathrm{min}$ before and after the transition from dim red light to bright white light conditions (\textit{Materials and Methods}, Movie S1). Before the onset of the light stimulus, the dim light adapted chloroplasts are found to be positioned in a single layer at the bottom of the cell as confirmed by confocal laser scanning microscopy (Fig.~\ref{fig:1}c). Within the first $20\,\mathrm{min}$ chloroplasts do not move far from their initial position, as indicated by the maximal travelling distance in Figure~\ref{fig:lightstim}a. As the light conditions are changed, the chloroplasts rapidly leave their local position and move far distances, corresponding to multiple chloroplast diameters.\\
Simultaneously, the instantaneous speed of the chloroplasts increases significantly by an order of magnitude (Fig.~\ref{fig:lightstim}b). To further study the difference in the motion in both regimes, we consider the mean squared displacement (MSD) (Fig.~\ref{fig:lightstim}c). Notably, the dynamics do not only show a higher baseline according to an anomalous diffusion coefficient $D$, but also a change of scaling exponent $\alpha$ from a mostly sub-diffusive ($\alpha<1$) to a super-diffusive ($\alpha>1$) scaling, when fitted with a powerlaw $MSD(\Delta t) = 4D \Delta t^\alpha$ at times $\Delta t=10-100\,\mathrm{s}$.\\ 
\indent We further quantify the motion of the chloroplasts, by defining a self-overlap function $Q(t)=\langle H(r_{th}-r(t))\rangle$ with $H(x)$ a Heaviside function $H(x)=1$ if $x\geq 0$ and $H(x)=0$ else, and $\langle\dots \rangle$ being the average over all particles. This function represents the temporal evolution of the fraction of chloroplasts which move not further than $r_{th}=5\,\mathrm{\mu m}$ (about one chloroplast diameter, see Fig.~\ref{SI_fig:Data}g) away from their initial position. The slow relaxation of this function at dim light conditions is due to smaller re-arrangements in the dim light adapted phase. Upon initiation of the white light stimulus, however, $Q$ drastically decreases, indicating the fast re-arrangements (Fig.~\ref{fig:lightstim}d).
This strong-light adaptation happens within a short period of $40\,\mathrm{min}$ and results in a cellular structure, which increases the transmittance of the leaf, hence decreases overall light absorption (Movie S1). We further quantified this transmittance by analyzing the raw intensity values from the microscope time series. We took the pixel intensity average $I(t)$ and calculated $\Delta I(t) = (I_{max}-I(t))/(I_{max}-I_{min})$ the relative difference to the maximal transmitted light intensity normalized by the difference between the minimal intensity after the onset of light and maximal transmitted intensity, when chloroplasts are aggregated (Fig.~\ref{fig:lightstim}d). This decrease in absorbance will cause smaller photo-damage, as has been studied previously in a variety of plants~\cite{Kasahara2002,Takahashi2011,Park1996,Li2009}.\\
\indent This transition from a dim- to a strong-light adapted state, is therefore of high physiological relevance for the plant. 
But how do chloroplasts behave under dim-light before the onset of the transition? How stable is the densely packed configuration?
These questions we aim to answer in the following.

\begin{figure}[h!]
		\centering
		\includegraphics[width=.85\textwidth]{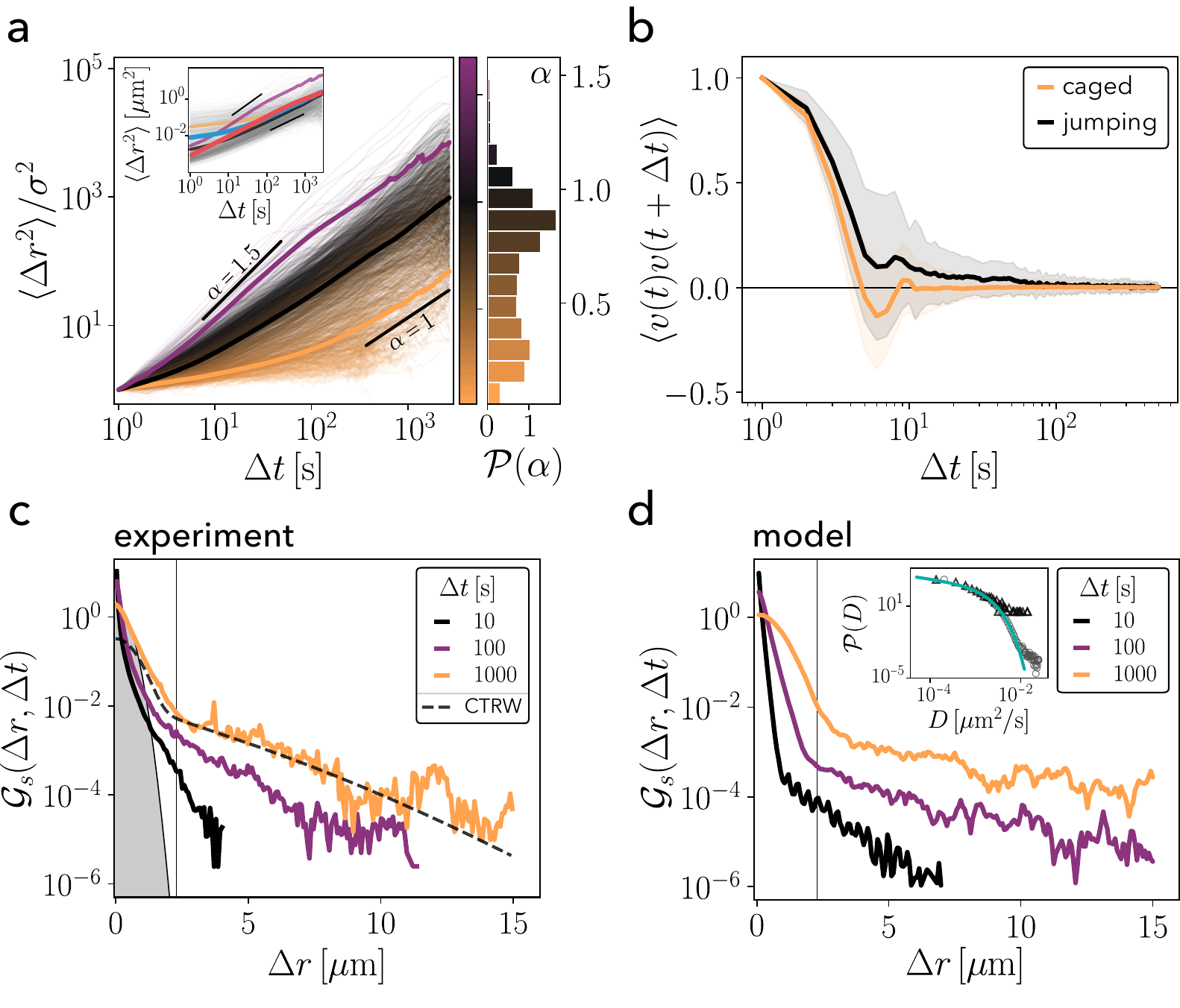}
		\caption{Trajectory analysis reveals sub- to super-diffusive transport and local trapping. \textbf{(a)} MSD of all trajectories relative to a minimal displacement scale $\sigma = MSD(\Delta t=1\,\mathrm{s})$. Colormap represents the power-law scaling exponent $\alpha$ fitted on short time scales below $100\,\mathrm{s}$. The exponent is widely distributed but peaks around a diffusive scaling regime ($\bar{\alpha} \approx 0.9$). While many trajectories show sub-diffusive scaling laws at small times, a few trajectories exhibit super-diffusive scaling $\Delta t \lesssim 100\,\mathrm{s}$. Thick lines: ensemble averages over super-diffusive trajectories $\alpha \geq 1.1$ (purple), diffusive trajectories $ 0.5< \alpha < 1.1 $ (black) and strongly sub-diffusive trajectories $\alpha \leq 0.5$. Notably on large time scales all regimes approach a diffusive scaling. Inset: MSD with units, most trajectories reach one micron displacement after long time ($t > 100\,\mathrm{s}$) only. Thick lines compare the ensemble average of all trajectories (blue line) to the ensemble averaged MSD of simulations (red line). Purple, black and orange lines correspond to ensemble averages in the main panel. Histogram: Bimodal distribution of scaling exponents $\mathcal{P}(\alpha)$ \textbf{(b)} Velocity autocorrelation functions (VACF) of trapped (orange) and jumping (black) chloroplasts. Some jumping chloroplast exhibit pronounced temporal correlation, which shifts the VACF to higher values. Anti-correlation as a sign of fast re-orientations is less pronounced in jumping chloroplasts, indicating a directed motion. \textbf{(c,d)} Self part of radial van Hove function at different lag-times for experiments (c) and model (d) reveals non-Gaussian displacements on short lag-times (compare with the gray area, representing a Gaussian function). For larger lag-times, the distributions are approaching a Gaussian at small displacements, hence the comparison with a continuous time random walk (CTRW) model~\cite{Chaudhuri2007} (dashed line in (c)) improves. The proposed model (d) accounts for the exponential tails, starting at a critical displacement (vertical line), and the exponential-to-Gaussian crossover at small displacements. Inset: diffusion coefficients of experiments (triangles) and simulation (circles) can be well described by a corrected exponential distribution (solid line).}\label{fig:2}
\end{figure}

\section*{Glassy Features of Chloroplast Motion under dim Light}
We track the positions of $1529$ chloroplasts under dim light and constant environmental conditions and segment their morphological features (\textit{Materials and Methods}). 
In dim light chloroplasts decrease their mobility~\cite{Haupt1982} and settle onto the periclinal walls in a single layer (Fig.~\ref{fig:1}c, Movie S1-S3). Chloroplast motion is based on the polymerization of specialized short actin filaments in the space between the chloroplasts' outer membrane and the cell membrane~\cite{Kadota2009,Kong2013} (Fig.~\ref{SI_fig:SchematicBio}). Therefore, membrane anchoring is essential for the propulsion of the organelle~\cite{Sakai2017}, which only allows for essentially two-dimensional motion.
Under dim light conditions, the binding probability of actin on the chloroplast is not biased in any direction, which will cause zero net motion. However, fluctuations, driven by random actin binding events, constitute a non-thermal random motion.\\
Chloroplasts appear as flat circular disks, with an average aspect ratio $AR=1.2$ and mean radius $R=2.29\,\pm\,0.29\,\mathrm{\mu m}$ (mean $\pm$ s.d.) corresponding to a polydispersity of about $12.6\,\%$ (Fig.~\ref{SI_fig:Data}g,j).
We measure a two-dimensional average packing fraction of approximately $73\,\%$, which is consistent with measurements in land plants~\cite{Honda1971,Ellis1985}. This suggests that chloroplast motion may depend on their close-by neighbors.\\
\indent Because they can move into all directions without rotating~\citep{Tsuboi2011}, the orientation of the chloroplast is uncoupled from the migration direction. Consequently, an analysis of the two-dimensional center of mass trajectories alone is sufficient to elucidate underlying stochastic processes.
The mean-squared displacement (Fig.~\ref{fig:2}a) reveals a broad spectrum of power law exponents $\alpha$ at small time scales ($\Delta t = 10-100\,\mathrm{s}$), ranging from sub-diffusive to super-diffusive trajectories (Histogram in Fig.~\ref{fig:2}a). At long time scales of $\Delta t \gtrsim 5\,\mathrm{min}$ the MSD approaches overall diffusive scaling ($\alpha=1$), with approximately exponentially distributed diffusion coefficients (Inset in Fig.~\ref{fig:2}d) and average diffusion coefficient $D\approx 8 \times 10^{-4} \,\mathrm{\mu m^2/s}$.
This diffusion coefficient is based on actin binding and polymerization~\cite{Kadota2009}, rather than thermal diffusion. To test this, we inhibited the active motion by a strong blue light shock, which drastically decreases the agitation, leaving the thermal diffusion undetectable (Fig.~\ref{SI_fig:Lightshock}, Movie~S5).\\
\indent The velocity auto-correlation reveals that the  velocity $v$ of chloroplast motion de-correlates within several seconds (Fig.~\ref{fig:2}b), indicating a persistent motion on these time scales. Together with the MSD scaling exponents, the velocity auto-correlation demonstrates a clear picture on the individual dynamics of chloroplasts: they are driven by a stochastic process which has a back-reflection effect, as suggested by the anti-correlation of the velocity after a few seconds (Fig.~\ref{fig:2}b). A few trajectories show slightly more persistent motion and larger scaling exponents indicating coordinated movement of actively driven chloroplasts.\\
\indent To further investigate the nature of the stochastic process constituting the motion of densely packed chloroplasts, we study the self-part of the van Hove function (step size distribution) $\mathcal{G}_s(\Delta r, \Delta t)$ (Fig.~\ref{fig:2}c) at various delay times $\Delta t = 10, 100 ,1000\,\mathrm{s}$. Strikingly, the van Hove function deviates from a Gaussian function at small and large displacements in two distinct manners. Firstly, small displacements are governed by a transition from a  non-Gaussian to a Gaussian step size distribution for increasing delay times. This feature is commonly found in the dynamics of (fractional) Brownian, non-Gaussian diffusive motion and has been observed in organelle and RNA transport~\cite{Wang2009,Lampo2017}. Secondly, the chloroplast step size distribution exhibits an exponential tail at all time scales.
The transition towards the exponential tail happens around $\Delta r \approx 2.3\,\mathrm{\mu m}$, corresponding to one average chloroplast radius $R\approx 2.29\,\mathrm{\mu m}$ (Fig.~\ref{SI_fig:Data}g).
This displacement is more than five times as large as the inter-chloroplast distance $l \approx 0.42\,\mathrm{\mu m}$, which is calculated via the difference of average chloroplast diameter and nearest neighbor distance~\cite{Debets2021} (Fig.~\ref{SI_fig:Data}g,f).
Hence, small displacements $\Delta r < l$ correspond to free motion until a chloroplast reaches $l$. Larger displacements $l \leq \Delta r < 2.3\,\mathrm{\mu m}$ are likely connected to the interaction of multiple chloroplasts, while displacements exceeding $2.3\,\mathrm{\mu m}$ show a sudden transition towards an exponential step size distribution.
Such an exponential tail has been previously observed in other systems on various scales, such as the bacterial cytoplasm~\cite{Parry2014}, particle transport in human cells~\cite{Aberg2021}, motion of cells in confluent tissues~\cite{Angelini2011} and non-active systems like super-cooled liquids~\cite{Ediger2000} and colloidal glasses~\cite{Kegel2000}. Using a continuous-time random walk (CTRW) model it was shown that exponential tails of the van Hove function display an universal feature of particle motion close to the glass transition~\cite{Chaudhuri2007,Chaudhuri2008}, which can be fitted to our experimental results (dashed line in Fig.~\ref{fig:2}c). However, this model does not fully account for the non-Gaussian center of the observed distribution, as we will discuss in more detail later.\\
\indent With a high packing fraction of $\phi \approx 73\,\%$ (\textit{Materials and Methods}) the amorphous chloroplast mono-layer bears a noticeable structural resemblance with dense colloidal systems, which may explain the key features of chloroplast motion. 
The direct study of individual trajectories in colloidal glasses has led to the conclusion that the stretched exponential step size distribution arises as a result of heterogeneous dynamics of slow and fast particles~\cite{Kegel2000}. This heterogeneity was found in active systems and also constitutes a hallmark for glasses~\cite{Janssen2019a}. Hence, our observation of the exponentially tailed step size distribution indicates the existence of dynamical heterogeneity underlying chloroplast motion.
\section*{Dynamic Heterogeneity due to the Co-existence of locally entrapped and re-arranging Chloroplasts}
\begin{figure}[htbp]
    \centering
    \includegraphics[width=1\textwidth]{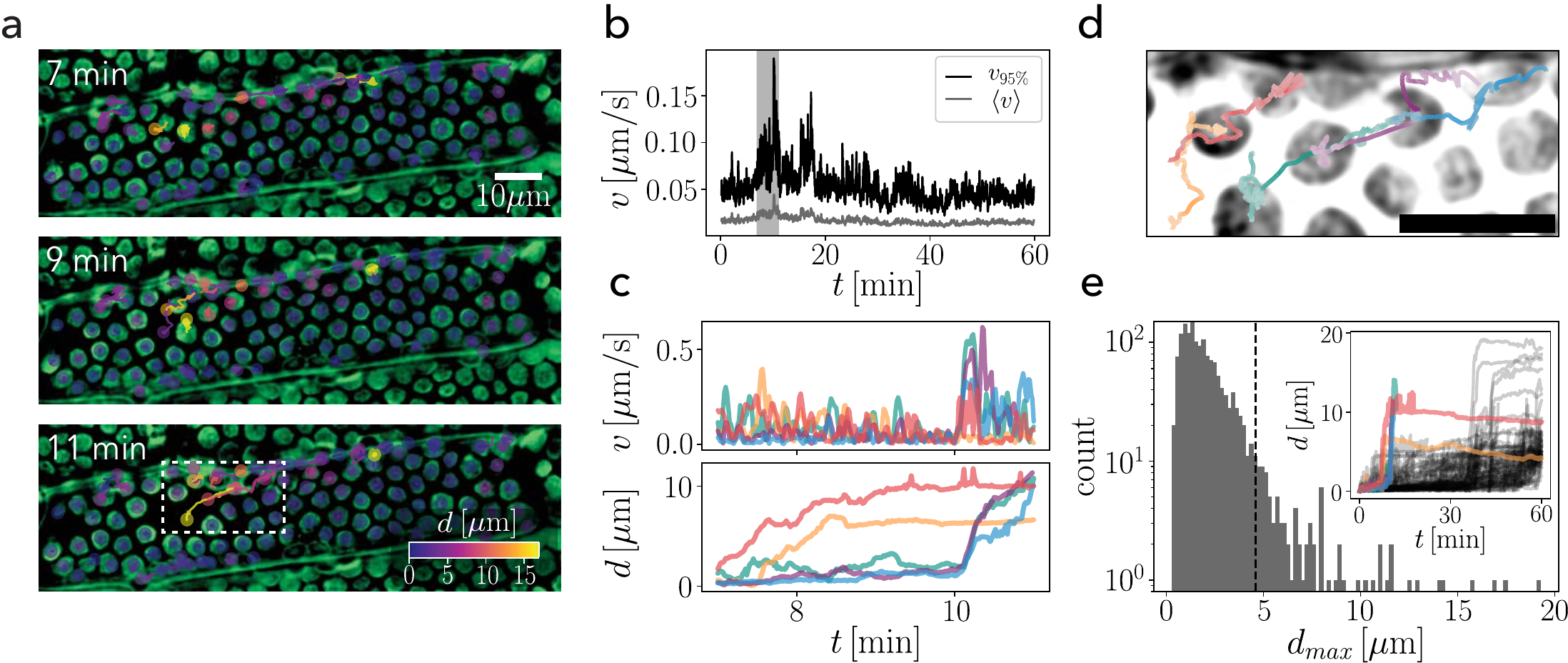}
    \caption{Intermittent re-arrangements of chloroplasts. \textbf{(a)} Time series of chloroplast motion during a bursting event. Trajectories are color-coded by their final distance $d$. While most chloroplasts remain at their initial location, a small region shows enhanced activity (white box). \textbf{(b)} Time series of average velocity (gray) and the $95$\textsuperscript{th} percentile velocity (black) of the cell in (a). The gray region around the peak corresponds to the time series in (a) and (c). The large increase of the $95$\textsuperscript{th} percentile indicates that the single event is an extreme of the velocity distribution. \textbf{(c)} Velocity and distance $d$ from initial position for trajectories which undergo strong re-arrangement (a, white box). Correlated motion of the sudden re-location becomes apparent. Colors for comparison of trajectories in (c) and (d). \textbf{(d)} Close-up of trajectories in (a), white box. Chloroplasts show random trapped motion, followed by transient collective motion, and subsequent fluctuations. Scale bar: $10\,\mathrm{\mu m}$. Higher color saturation along the trajectories indicates enhanced speed. \textbf{(e)} Histogram of maximal displacements with threshold value (dashed line, \textit{Materials and Methods}). Inset: Distance from initial position for all chloroplasts exceeding the distance threshold showing similar pattern: sudden re-arrangements and subsequent plateau at approximately discrete positions of a few chloroplast diameters. Trajectories from (c) and (d) are marked in respective colors.}
    \label{fig:3}
\end{figure}

The aforementioned hypothesis requires analysis of individual chloroplast trajectories. 
We find qualitatively different trajectories. While many chloroplasts remain caged by their neighbors (slow particles), a small fraction ($88$ out of $1529$) can escape their local confinement and move for longer distances until they get trapped again as a jump-like motion (fast particles) (Fig.~\ref{fig:3}a,e). 
The dynamic heterogeneity is studied by disentangling the fast and slow dynamics of the system, based on the chloroplasts maximal displacement $d_{max}$ during the experimental time (Fig.~\ref{fig:3}e)~\cite{Kob1997,Kegel2000} and their maximal speeds (see \textit{Materials and Methods}).\\
\indent We uncover large regions of trapped particles and smaller regions of more mobile chloroplasts (Fig.~\ref{fig:3}a, Fig.~\ref{SI_fig:Data}a) by marking trajectories by their total displacement from their initial position. Fast chloroplasts re-arrange within a short period of time and arrest the dynamics afterwards.
The intermittent bursts of chloroplast re-locations can be detected in the tails of the speed distribution (Fig.~\ref{fig:3}b), as represented by the $95$\textsuperscript{th} percentile, while the average speed does not exhibit strong changes. This confirms that re-arrangements are connected to exceptionally high speeds of chloroplast motion, which transiently occur in a small fraction of all chloroplasts in a cell.
Besides the spatial co-localization of chloroplast re-locations, we find that their motion can be temporally correlated and follows a string-like profile. Individual chloroplasts can break out of their local cage and push their neighbors with them (Fig.~\ref{fig:3}c,d, Fig.~\ref{SI_fig:Data}a).
Additionally, fast chloroplasts have longer velocity auto-correlation times, suggesting an increase in the persistent motion, while slow particles oppose their motion on smaller time scales (Fig.~\ref{fig:2}b).
The movement of all re-arranging chloroplasts follows a similar scheme: a sudden chloroplast displacement at elevated velocities plateaus at a distance of only a few chloroplast diameters (Fig.~\ref{fig:3}e inset). This suggests that they are locally entrapped again.\\
\indent Differences in the dynamics, however, cannot be simply ascribed to structural differences as suggested by the absence of strong correlations between structural features like the particle radius, average nearest neighbor distance and aspect ratio and dynamic features like the maximal travelling distances and maximal speeds of individual trajectories (Fig.~\ref{SI_fig:Data}a-e).
Similar sudden and hardly predictable re-arrangements have been found in simulations of active Ornstein-Uhlenbeck particles~\cite{Mandal2020}, sheared granular media~\cite{Marty2005} and in epithelial tissues~\cite{Angelini2011}. Interestingly, the string-like cooperative motion of particles in a close-to glassy state were observed in super-cooled liquids~\cite{Cicerone2014,Caporaletti2021}, simulations of Lennard-Jones mixtures~\cite{Donati1998} and granular beads~\cite{Keys2007}. \\
\indent The striking similarity to many systems close to the glass transition implies that chloroplast motion in this dense single-layered configuration obeys similar physical constraints and mechanisms.
In the following, we will build a mathematical formulation based on physiological parameters that models the co-existence of trapped particles and regions of higher mobility in the chloroplast dynamics. This approach allows us to study the effect of activity and crowding on the chloroplast dynamics. Our goal is to infer the stability around the critical state of the chloroplast mono-layer.
\section*{A Threshold-Based Jump Diffusion Model reveals that the dim light adapted Chloroplast mono-layer is close to a Glass-transition}
\begin{wrapfigure}{r}{0.5\textwidth}
	\centering  
	\includegraphics[width=0.48\textwidth]{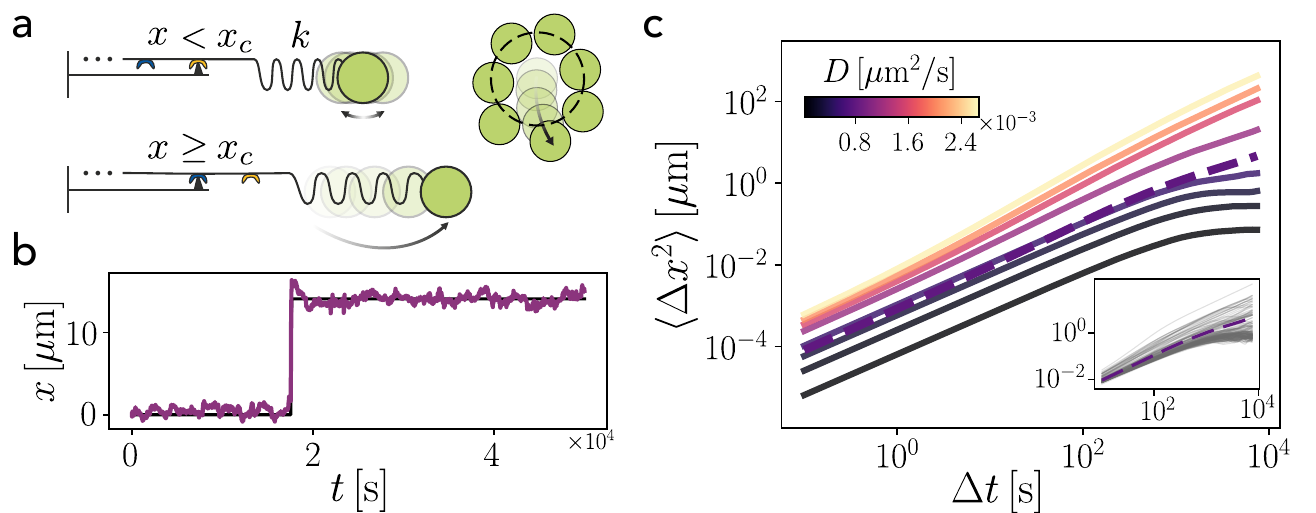}
	\caption{ Model based analysis suggests the vicinity of the chloroplast system to a dynamical transition. \textbf{(a)} Sketch of the mechanistic model. A Brownian particle is connected to a spring and a jump-element which undergoes sudden extension or compression if a critical displacement is reached. \textbf{(b)} Example trajectory shows sudden jumping events before being trapped again. \textbf{(c)} Ensemble-averaged MSD from low to large steady-state diffusion coefficients $D$ shows a transition from trapped to free motion. Parameters obtained from our data suggest that the system is close to a transition towards unconfined motion. Inset: variability of trajectories shows dynamics ranging from caged to jumping motion, dashed line: ensemble average MSD with parameters from experiment.}
	\label{fig:4}
\end{wrapfigure}

We propose a stochastic model of chloroplasts to elucidate the impact of critical length-scales and active diffusion of the system. In this model discrete jumps are triggered by a threshold-based mechanism (Fig.~\ref{fig:4}a).
We consider a particle in a harmonic potential $V(x) = -\frac{k}{2} x^2$ that can undertake a jump of finite velocity $v$ upon exceeding a critical distance $x_c$ from the potential minimum position $a(t)$. This leads us to a Langevin-equation:
\begin{equation}
\label{eq:1}
   \dot x = \begin{cases} -\frac{x}{\theta}+\sqrt{2D(t)} \xi(t)&, \; |x-a(t)|<x_c\\
-\frac{x_c}{\theta} \pm v + \sqrt{2D(t)} \xi(t)&,\;|x-a(t)| \geq x_c
\end{cases}
\end{equation}
where $\xi(t)$ is a Gaussian white noise $\langle\xi \rangle = 0$, $\langle \xi(t)\xi(t')\rangle = \delta (t-t')$.
The effective spring constant $k$ represents a confinement by actin-mediated anchoring~\cite{Sakai2017} and the neighboring particles~\cite{Doliwa1998,Weeks2002}. Together with a damping pre-factor $\gamma$ it defines a correlation time $\theta=\frac{\gamma}{k}$. Our analysis only allows us to estimate the time scale $\theta$ from spatial auto-correlations of the trajectories, but the exact physical interpretation of $\gamma$ and $k$ in the experiments is not possible.
After every jump the center position of the potential $a$ is translated by an equal amount to maintain the particle inside the harmonic potential.\\ 
\indent In our experiments, we observe a non-Gaussian to Gaussian transition over time (Fig.~\ref{fig:2}c). Such transition is likely caused by the underlying process of chloroplast motion, which is based on actin polymerization~\cite{Kadota2009} in a highly dynamic and heterogeneous environment, resulting in temporal fluctuations of the active diffusion coefficient $D$. Therefore, we model the stochastic dynamics of the diffusion coefficient $D(t)$ by the square of an Ornstein-Uhlenbeck process~\cite{Chubynsky2014,Chechkin2017}:
In the equation above, $\tau$ is a relaxation time scale, and $\sigma$ is the noise amplitude of a Gaussian white noise $\eta(t)$. Similar formulation like in \eqref{eq:2} has been used previously to describe active processes in cytoskeletal dynamics~\cite{Toyota2011} 
and the transport of RNA~\cite{Lampo2017} or colloidal beads on membranes~\cite{Wang2009}. In the present context, \eqref{eq:2} models fluctuations of the driving mechanism and environmental heterogeneity and leads to temporal correlations of the active diffusion coefficient. We support this assumption by observing that this ``diffusing diffusivity'' model~\cite{Chubynsky2014} predicts an exponential distribution of diffusion coefficients~\cite{Chechkin2017} consistent with measurements of the diffusion coefficient of caged particles at times beyond $5\,\mathrm{min}$ (Fig.~\ref{fig:2}d, inset). This allows us to use the average active diffusion coefficient $\langle D \rangle = \sigma^2 \tau = 8 \times 10^{-4} \,\mathrm{\frac{\mu m^2}{s}}$ to estimate parameters for the model. Other parameters are extracted from particle trajectories and physiological parameters (\textit{Materials and Methods}). A typical trajectory of this model displays diffusive dynamics with time-correlated variance and sudden jumps of discrete size (Fig.~\ref{fig:4}b). The resemblance to the  experiments is also reflected in the van Hove function (Fig.~\ref{fig:2}c).\\
\indent The discrete jumps can account for the exponential tails, while the
non-Gaussian to Gaussian transition for increasing timescales and small displacements in the van Hove function can be well described by the diffusing diffusivity model.
\indent Note that our model neglects the subdiffusive dynamics at small time scales (Fig.~\ref{fig:2}a) by construction~\cite{Chubynsky2014,Chechkin2017}. Together with the damping parameter and the critical length scale, the effective diffusivity $D=\sigma^2 \tau$ steers the caging dynamics, as observed in the MSD of the trajectories (Fig.~\ref{fig:4}c, Fig.~\ref{SI_fig:Model}a). At a small noise amplitude $\sigma$ the system is strongly caged, as chloroplasts are drastically less probable to reach a high enough displacement to jump. 
\begin{equation}
\label{eq:2}
    D(t)=Y(t)^2, \; \dot Y(t) = -\frac{Y}{\tau} + \sigma \eta(t).
\end{equation}
If, however, $\sigma$ is increased by $20-40\%$, the jumping dynamics are more pronounced, suggesting that chloroplasts are close to a critical point and can easily transition into a \textit{liquid-like} state. In our model this transition is, to a great extent, governed by the rate of jump events undertaken by the particle. We define the jump time distribution by construction  as the first passage time distribution of an Ornstein-Uhlenbeck process at long timescales, at which the fluctuating diffusion coefficient approaches $D = \sigma^2 \tau$. The first passage time distribution approaches a series of exponential functions~\cite{Ricciardi1988} and the mean first passage time is controlled by a dimensionless quantity $x_c/\sqrt{D\theta}$~\cite{Nobile1985}. This suggests that a similar transition is also possible when the critical length-scale $x_c$ increases, while $D$ is held constant~(Fig.~\ref{SI_fig:Model}), which has been observed in studies of plant cells with varying chloroplast numbers and areas~\cite{Koniger2008}. Hence, despite the simplified description of the cage effect and jumping dynamics, this model is sufficient to predict that, not only the activity $\sigma$, but also the critical size $x_c$ can easily drive the system out of the caged dynamics.
\section*{Conclusion}
Plants have developed multiple adaptation mechanisms to changing light conditions~\cite{Li2009}. One of these mechanisms is orchestrated via independent movements of disk-like chloroplasts towards or away from light.
The motion of densely packed chloroplasts adapted to low light conditions exhibits a striking similarity to the caged dynamics of super-cooled liquids or colloidal suspensions close to glass transition. We observe actively driven organelles confined in a single layer at high packing fraction. Intermittent coordinated motion of single or multiple chloroplasts leads to a exponentially-tailed step size distribution caused by the heterogeneous nature of the environment. Furthermore, we uncover dynamical features similar to systems close to the glass transition, such as dynamically heterogeneous regions, sudden chloroplast displacements and even string-like coordinated movements. This renders chloroplasts a novel biological system exhibiting glassy behavior, while maintaining an unaltered disk-like shape, contrary to other biological active glasses like epithelial cell sheets~\cite{Angelini2011,Garcia2015,Sadati2013,Atia2018,Mongera2018,Park2015,Oswald2017}. A few other examples exist in which dense colloidal-like active particles play a physiological role, for example in the context of active fluidization of statoliths, granular organelles in gravisensing plant cells~\cite{Berut2018}, or the involvement of nuclear crowding in tissue stabilization~\cite{Kim2022}.\\
\indent We compare the observed dynamics to a one-dimensional model, whose parameters are obtained from experiments and physiological values. The statistics of this model are in agreement with our data. This indicates that burst-like chloroplast jumps could be triggered upon exceeding a threshold, similar to the cage length in glasses~\cite{Debets2021}. By construction, the waiting times between the discrete jump events in this model are identical to the first-passage time of an Ornstein-Uhlenbeck process with diffusing diffusivity, which is approximately exponentially distributed for large enough boundaries and times~\cite{Ricciardi1988} (Fig.~\ref{SI_fig:Model}c). This indicates that the underlying statistics of this model are similar to those of the continuous time random walk (CTRW) model proposed by Chaudhuri et al.~\cite{Chaudhuri2007,Chaudhuri2008}, which is used to describe a universal step size distribution function in various systems close to the glass transition.
However, this CTRW model needs the mean waiting time between jump events as a fit parameter, which is difficult to infer from limited data directly. Our model enabled us to convert a waiting time between jump events into a length scale $x_c$, which can be more easily extracted from data. 
Varying the model parameters suggests that chloroplasts operate at a meta-stable point and can easily perturb the mono-layer structure by slight enhancements of their activity, which could explain the fast adaptation mechanism towards strong light, as observed in our experiments (Fig.~\ref{fig:lightstim}).
The model also hints towards the important role of a cage effect inhibiting chloroplast photo-relocation motion when the chloroplast size is changed, as has been observed before~\cite{McCain1998,Koniger2008}.\\
\indent We interpret our results as follows: chloroplasts accumulate into a single layer in response to weak light in order to achieve a high surface coverage and therefore maximize photosynthetic efficiency~\cite{Gotoh2018}. Since we know that chloroplasts avoid highly intense light~\cite{Wada2003}, they must maintain the ability to undertake this avoidance motion efficiently without being hindered by other chloroplasts. Therefore, the ability to increase the activity and a sufficiently loose packing of chloroplasts are necessary. Our study shows that chloroplasts in dim light conditions resemble a system close to the glass transition. 
Such a dynamical transition depends on the microscopic features of the organelles propulsion mechanism, as well as its cellular surrounding and can deviate from standard model glass forming systems. For example, chloroplasts can create protein bridges~\cite{Kong2020} and actin-cages~\cite{Kadota2009}, hence, the systems behavior may differ from purely repulsive systems~\cite{Zaccarelli2009}.\\
\indent How chloroplasts spread on the periclinal cell walls as a result of an accumulation response under dim light remains a subject for further research, since a direct observation of this process was not possible. Further studies on collective escape from this low-light adapted state towards an avoidance configuration will shine light on the various dynamic phase transitions this system can undergo, which were also found in dense active matter systems~\cite{Mandal2020,Keta2022}.
Additionally, unveiling the exact interactive forces between the organelles and the impact of confinement due to the cell walls on the transitions between dark and light-adapted states remains a matter for further investigation.

\section*{Materials and Methods} \label{note:Note1} 
\subsection*{Preparation and Imaging}
\textit{Elodea densa} was kept in an aquatic culture. Ambient light conditions were applied $1\,\mathrm{h}$ before image acquisition. 
For imaging, a low-light adapted leaf was detached and placed between two glass slides. We study the lower cell layer of the two-layered leaf tissue~\cite{Rascio1991}, as these are more strongly involved in photosynthesis~\cite{Maai2020}. The upper epidermal cell layer is important to enhance gas exchange due to reduced gas diffusion in water~\cite{Pedersen2013}.
Bright field microscopy was performed with a \texttt{Nikon TI2} microscope using a halogen light source and a red-light $620\,\mathrm{nm}$ cut-on wavelength filter at low light irradiance of approximately $1\,\mathrm{W/m^2}$ in combination with a \texttt{Photometrics BSI Express} sCMOS camera with high quantum yield to enable imaging every $1\,\mathrm{s}$ with a $60\times$ water-immersion objective (NA$=1.2$) and pixel resolution of $0.11\,\mathrm{\mu m/px}$. High light intensities for avoidance motion are achieved by removing the red-light filter and increasing the irradiance to approximately $600\,\mathrm{W/m^2}$, approximately corresponding to the solar irradiance on a sunny day.
The strong light shock was performed using the fluorescence lamp (\texttt{CoolLED pE-300}) at a peak wavelength of $\lambda=466\,\mathrm{nm}$ with an irradiance of approximately $10\,\mathrm{kW/m^2}$ inducing a rapid and irreversible de-polymerization of actin and fixes of the sample.
Confocal laser scanning microscopy was performed using a \texttt{Leica SP8}. The $638\,\mathrm{nm}$ excitation line and a broad band emission filter $650-750\,\mathrm{nm}$ were used to image chlorophyll autofluorescence. Z stacks were acquired over heights of $20\,\mathrm{\mu m}$ in $0.5\,\mathrm{\mu m}$ steps, while the z-resolution is $860\,\mathrm{nm}$ with an $60\times$ oil immersion objective ($NA=1.4$).
\subsection*{Image Processing}
Image time series were background subtracted using a difference-of-Gaussian method with variances $\sigma_1 = 2\,\mathrm{px}$ and $\sigma_2=37\,\mathrm{px}$ and subsequently inverted (Fig.~\ref{SI_fig:Methods}). The resulting image shows white circular spots, reminiscent of nuclei training data for the StarDist versatile model~\cite{Weigert2020}. Using this pre-trained StarDist network we segmented the images and detect center of mass positions of the particles. The segmented objects were analysed for their area and aspect ratio. Small objects of an area of $250\,\mathrm{px}$, corresponding to an equivalent diameter of less than $1.78\,\mathrm{\mu m}$, were removed. Additionally, elongated objects with aspect ratios of above $3.5$ or below $0.3$ were excluded to avoid mis-detections of cell walls. We hand-segmented the surrounding cell wall to discriminate chloroplasts in different cells (Fig.~\ref{SI_fig:Data}b). Histograms for the particle diameters and aspect ratios, as well as the packing fraction, were calculated based on hand-segmented chloroplasts, to avoid detection biases and smaller mistakes in the StarDist-based segmentation.
The average packing fraction was calculated by the sum of all areas of segmented chloroplast devided by the area of the segmented cells. 
\subsection*{Trajectory Analysis}
The center-of-mass positions of all segmented chloroplasts were linked to trajectories using a linear assignment problem solver \textit{trackpy}~\cite{trackpy}. Spurious trajectories of a duration smaller than $400\,\mathrm{s}$ were not considered in the analysis. Trajectories were classified as \textit{fast} when their maximal displacement lies above the mean particle size (dashed line in Fig.~\ref{fig:3}c, Fig.~\ref{SI_fig:Data}g) ~\cite{Kob1997,Kegel2000}, and whose maximal velocity lies above $0.1\,\mathrm{\mu m /s}$. Particle velocities were calculated with the first derivative of the third-order Savitzky-Golay filtered data, with a kernel length of $11\,\mathrm{s}$. The self-part of the van Hove function and the mean squared displacements were calculated with trackpy~\cite{trackpy}, whilst the MSD scaling exponents $\alpha$ were extracted using a local fitting scheme~\cite{Maier2012} between $\Delta t = 10-100\,\mathrm{s}$. As we track the center of mass of the segmented particles, we have two contributions to the error of the tracking algorithm: one from the limited pixel-size of $\sigma_{px} = 0.11\,\mathrm{\mu m}$ and area fluctuations of the mass. As the mask can differ from image to image, we take the area-fluctuations as a proxy to estimate a detection error: $\sigma \approx \sigma_{px}\sqrt{\sigma_A/\langle A \rangle}  \approx 0.07\,\mathrm{\mu m}$, with the mean $\langle A \rangle$ and the standard deviation $\sigma_A$ of the areas obtained by temporal and ensemble averaging over all trajectories. 
\subsection*{Simulations}
Our model (equations \eqref{eq:1}, \eqref{eq:2}) was integrated using a stochastic Euler-Maruyama scheme with a $10^6$ time steps of duration $dt=0.1\,\mathrm{s}$. Simulation parameters were retrieved from trajectories. We choose the $1/e$ decay of the positional auto-correlation function (Fig.~\ref{SI_fig:Data}i) to estimate the harmonic relaxation time $\theta=1063\,\mathrm{s}$.
Although the function does not follow an exponential decay it gives a reasonable estimate for the correlation time.
Additionally, we estimated the noise amplitude $\sigma$ of the diffusing diffusivity from the measurements of the long-time diffusion coefficients of the system as $D=\sigma^2 \tau$. Here we set the correlation time $\tau=90\,\mathrm{s}$ which is well within the range of actin turnover times~\cite{Blanchoin2010}.
The jump velocity $v\approx0.6\,\mathrm{\mu m/s}$ was estimated from the average maximal velocity of jumping chloroplasts. The critical length scale $x_c=2.3\,\mathrm{\mu m}$ is the point at which the van Hove function transitions to an exponential (Fig.~\ref{fig:2}c). This critical length scale is close to the average particle radius (Fig.~\ref{SI_fig:Data}g). For parameter sweeps, $x_c^{-1}$ and $\sigma$ were changed from $20\,\%$ to $180\,\%$ of their experimental value, and simulations were repeated $200$ times for each parameter.
Simulation data was analysed exactly the same as trajectories (see above). 
The continuous time random walk model (Fig.~\ref{fig:2}c, dashed line) was implemented as outlined in~\cite{Aberg2021} with parameters $\tau_1=3000\,\mathrm{s}$, $\tau_2 = 100\,\mathrm{s}$,  $d=0.6\,\mathrm{\mu m}$, $l=1.2\,\mathrm{\mu m}$.

\section*{Acknowledgements}
We are grateful to Federico Caporaletti, Vincent Debets, Liesbeth Janssen, Ludovic Berthier and Carola Seyfert for invaluable discussions and assistance. N.S. thanks Robert Haase for insightful discussions on object tracking at an early stage of the project.

\printbibliography

\clearpage
\renewcommand\thefigure{S\arabic{figure}}
\setcounter{figure}{0}    
\section*{Supplementary Information}
\subsection*{Supplementary Movies}
Supplementary movies S1-S5 can be found under this link: \url{https://drive.google.com/drive/folders/1ezeK4Iy6htm5bvcWWkMW2JPsMTKHkisz?usp=sharing}\\

\textbf{Movie S1}: Time lapse ($65\,\mathrm{min}$) of multiple \textit{Elodea densa} leaf-cells with chloroplasts, acquired with brightfield microscopy. At $t=20\,\mathrm{min}$ white light is turned on and chloroplasts break out from their initially two dimensional densely packed configuration into aggregates. These dynamics are a strong light avoidance response.\\
\textbf{Movie S2}: Time lapse ($10\,\mathrm{min}$) of Chloroplasts acquired with a laser scanning confocal microscope. The dense dim-light adapted configuration of the organelles is shown to be in a two dimensional layer. Caged motion within the local cages becomes apparent.\\
\textbf{Movie S3}: Time lapse ($60\,\mathrm{min}$) of background subtracted brightfield images of multiple cells filled with chloroplasts, false colors. Microscopy under low light assures that the system remains in a dim-light adapted state. Few re-arrangement motions are visible.\\
\textbf{Movie S4}:Close-up on a specific cell in Movie 4, showing strong correlated re-arrangement motions.\\
\textbf{Movie S5}:Time lapse ($60\,\mathrm{min}$) of bright field microscopy, after a strong light pulse is given. The light pulse saturates the chloroplasts and leads to a rapid de-polymerization of actin. Therefore no movement besides drift of the microscope is detectable. Hence, thermal motion of the chloroplasts is negligible.

\begin{figure}[h]
    \centering
    \includegraphics[width=0.95\textwidth]{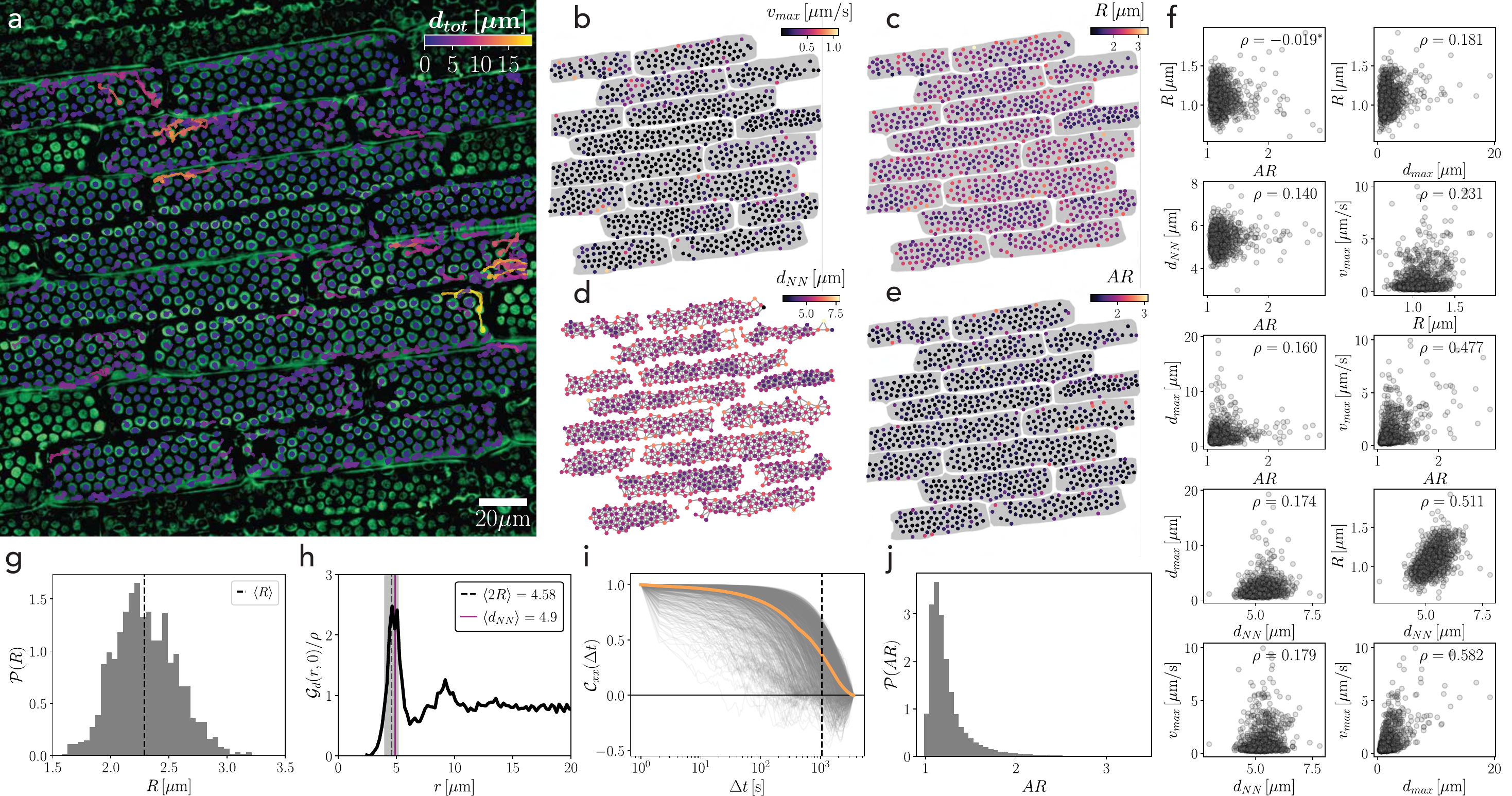}
    \caption{Specifications of particle trajectories. \textbf{(a)} All trajectories, colored by the total distance shows localized fast displacements $d_{tot}$. \textbf{(b)} Chloroplasts at time point $t=0$ colored by their maximal velocity during re-arrangements. Correlates with distance in (a). Gray mask: segmented cells. \textbf{(c)} Chloroplast positions colored by radius $R$. \textbf{(d)} Nearest-neighbor network (gray lines) of chloroplasts at $t=0$ colors correspond to average nearest neighbor distance $d_{NN}$. \textbf{(e)} Chloroplasts colored by aspect ratio. \textbf{(f)} Correlations of dynamical features: maximal distance $d_{max}$ and maximal velocity $v_{max}$ and structural features: radius $R$, aspect ratio $AR$ and local average nearest neighbor distance $d_{NN}$. Plots are sorted by correlation coefficient $\rho$. Insignificant correlations (significance level $R<0.01$) indicated with star. \textbf{(g)} Histogram of particle radii $R$ with the critical distance $x_c$ (dotted line) corresponding to the mean radius. \textbf{(h)} Radial distribution function of the particles around the average density $\rho$, with a nearest neighbor peak around two particle radii (dotted line) plus their standard deviation (gray). The nearest neighbor shell size (purple) can be estimated from an average until the first minimum. \textbf{(i)} The autocorrelation function of the position for all particles (gray) and the ensemble average (orange). The $1/e$ decay is located at $\Delta t=1063\,\mathrm{s}$ (dotted line). \textbf{(j)} Histogram of particle aspect ratios $AR$.}
    \label{SI_fig:Data}
\end{figure}
\newpage
\begin{figure}[htbp!]
    \centering
    \includegraphics[width=1\textwidth]{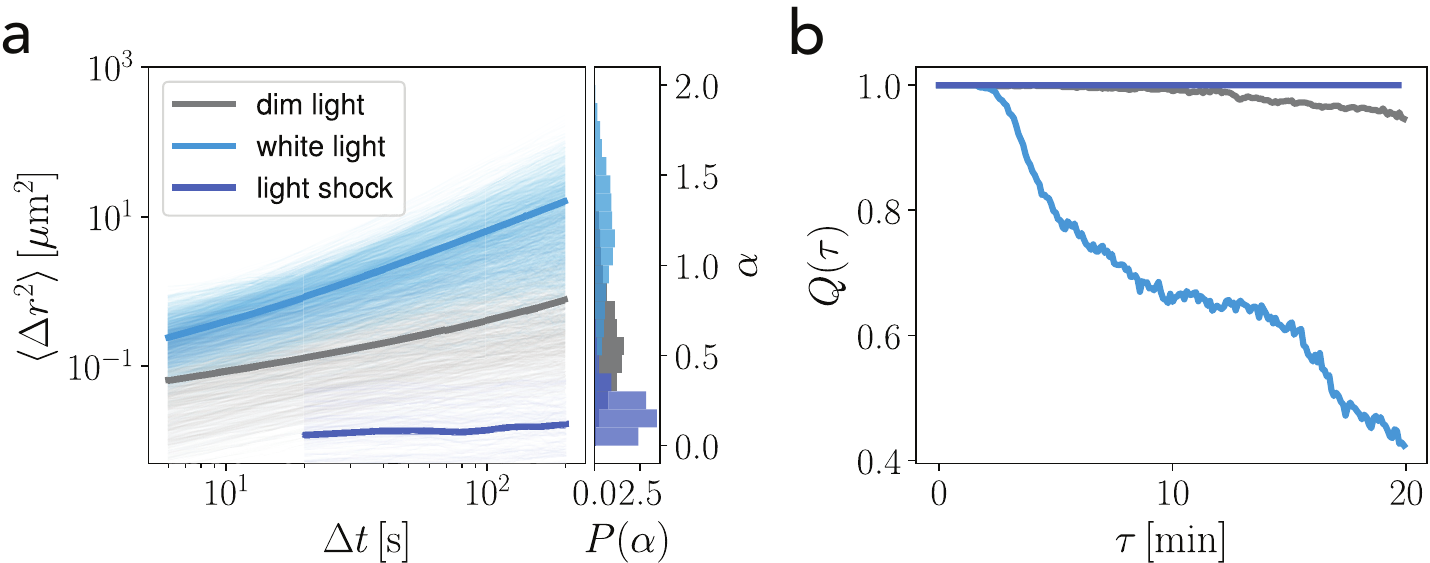}
    \caption{Experiment after light shock compared to dim and bright light phase. \textbf{(a)} Time averaged mean squared displacements for individual trajectories at dim light (gray), after white light stimulus (light blue) or after light shock experiment (dark blue) with ensemble averages as thick lines, respectively. No motion is visible for the light shock. The non-zero mean squared displacements are due to tracking errors and in the order of a single pixel. This regime highly differs from dim light adapted motion or the agitated motion during the bright light adaptation.
    Right: Probability distribution function for fitted power-law scaling exponents $\alpha$, showing strongly sub-diffusive motion ($\alpha\approx0.15$) after a strong light shock. This proves that motion at dim light is dominated by a-thermal noise. \textbf{(b)} Light blue and light gray line are the same as in Fig.~2d. After Light shock, overlap function (dark blue line) indicates no significant movement.}
    \label{SI_fig:Lightshock}
\end{figure}
\begin{figure}[h]
    \centering
    \includegraphics[width=1\textwidth]{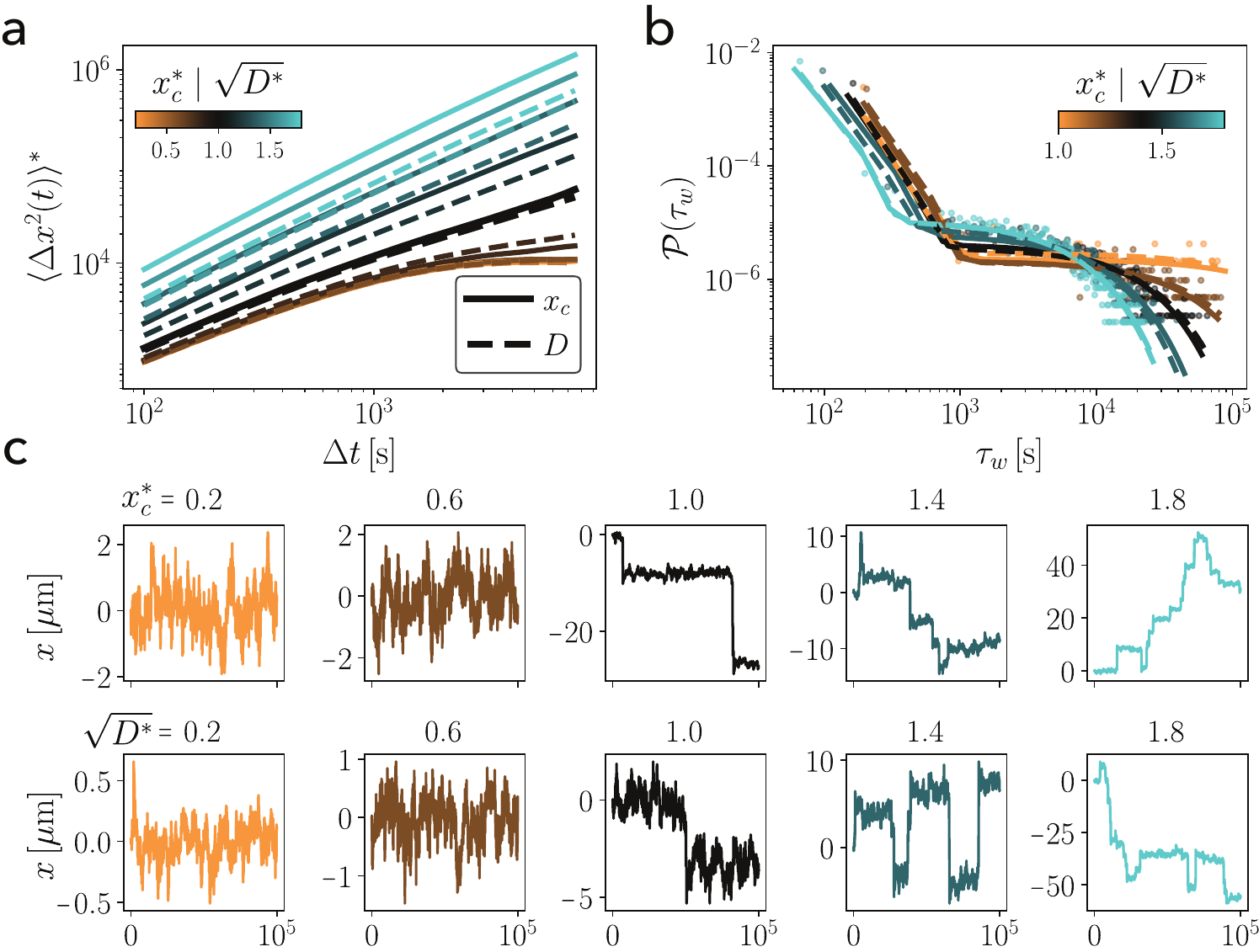}
    \caption{Parameter variations in the jump model. \textbf{(a)} Mean squared displacement, normalized by the displacement at $\Delta t=0$ for different critical length scales $x_c* = x_{c_0}/x_c$ (solid lines), where $x_{c_0}=2.3\,\mathrm{\mu m}$ is the experimental value. Additionally, a comparable parameter sweep is performed by changing the activity $\sigma \sim \sqrt{D^*}$ (dotted lines) with relative diffusion coefficients $D^*=D/D_0$, where $D_0= 8 \times 10^{-4}\,\mathrm{\mu m^2/s}$ is the experimental value. \textbf{(b)} Waiting time distributions for same parameters as in panel a (data points shown for $x_c^*$) with double-exponential fits for $x_c^*$ (solid lines) and $D^*$ (dashed lines). \textbf{(c)} Example trajectories for different $x_c^*$ (top) and $D^*$ (bottom), for larger $D^*$ or $x_c^{-1}$, more jump events lead to larger displacements resembling a transition towards a more \textit{liquid} state.}
    \label{SI_fig:Model}
\end{figure}
\begin{figure}[htbp!]
    \centering
    \includegraphics[width=0.95\textwidth]{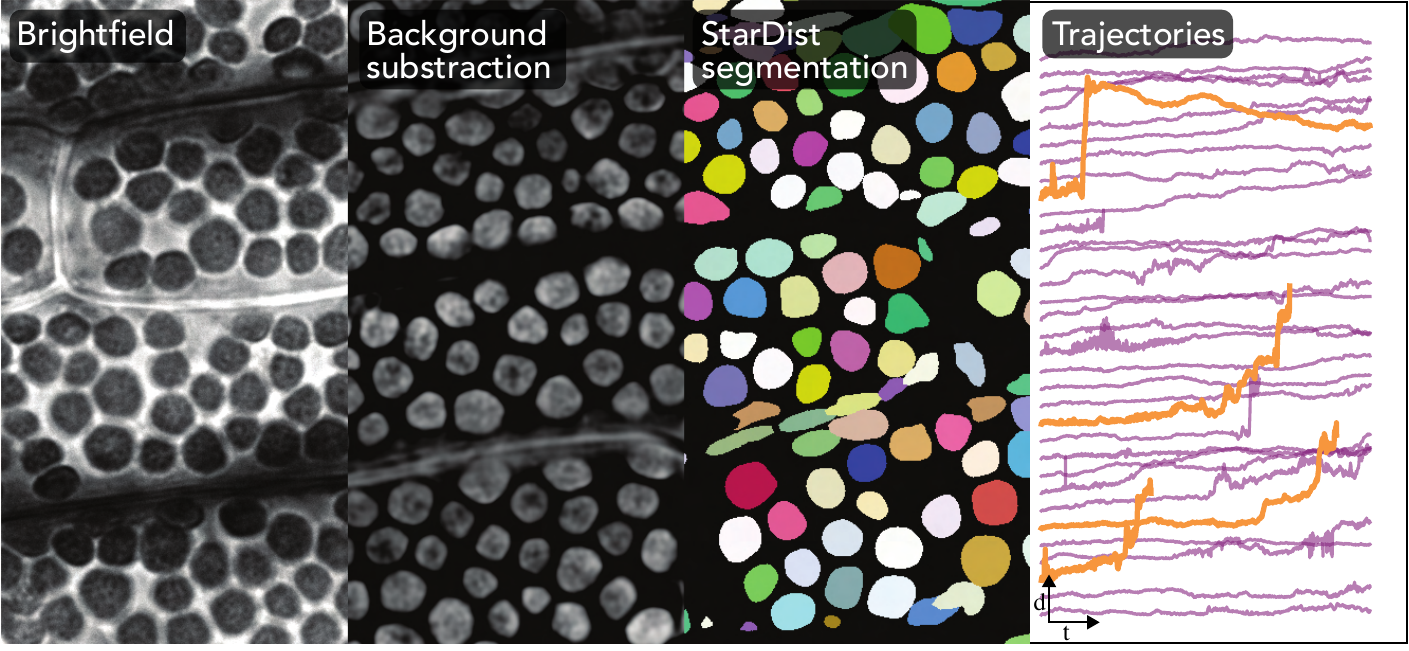}
    \caption{Image processing workflow: brightfield images (left) are background subtracted using a difference-of-Gaussian method. The subtracted image is segmented into individual regions using a StarDist neural network. Each region can be analyzed for its area, aspect ratio and center of mass position. This procedure is done for every frame of the image time series. The positions are assigned to trajectories using trackpy (right).}
    \label{SI_fig:Methods}
\end{figure}
\begin{figure}[htbp]
    \centering
    \includegraphics[width=0.5\textwidth]{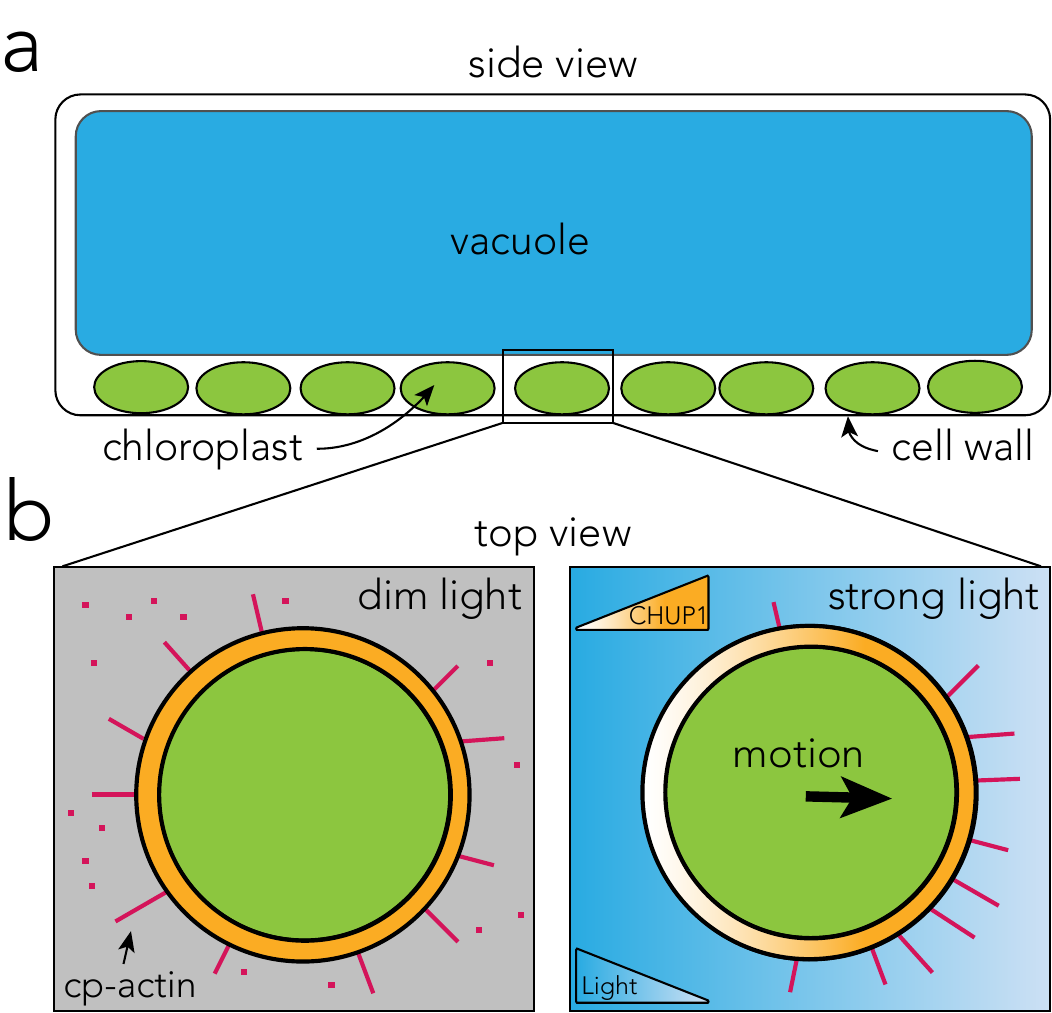}
    \caption{Cell architecture and driving mechanism of chloroplasts. Figure inspired by~\cite{Wada2018}. \textbf{(a)} Leaf cell structure of \textit{Elodea densa} after adaptation towards dim light. Disk-like chloroplasts are sitting on the lower plasma membrane, while light is coming from top. The vacuole is consuming most of the space of the cell. \textbf{(b)} Top view on a single chloroplast under different light conditions. Short actin (cp-actin, red bars) polymerization leads to chloroplast motion~\cite{Kadota2009,Kong2013}. Under homogeneous dim light conditions (left) cp-actin filaments are homogeneously distributed on the chloroplast surface. The actin polymerization leads to random force with zero average. In light gradients avoidance or accumulation motion leads to an effective drift by re-distributing actin binding domains (CHUP1)~\cite{Kong2020}. In-depth summary on anchorage and molecular propulsion mechanism can be found in \cite{Wada2016,Wada2018}.
    }
    \label{SI_fig:SchematicBio}
\end{figure}

\end{document}